\documentclass{ws-procs9x6}

\newcommand{\BR}{{\cal B}}

\newcommand{\piz}{\pi^0}

\newcommand{\psp}{\psi(2S)}

\newcommand{\jpsi}{J/\psi}

\newcommand{\pp}{\pi^+\pi^-}

\newcommand{\beq}{\begin{equation}}
\newcommand{\eeq}{\end{equation}}
\newcommand{\bitm}{\begin{itemize}}
\newcommand{\eitm}{\end{itemize}}
\newcommand{\op}{\omega\phi}
\newcommand{\oo}{\omega\omega}
\begin{document}

\title{Search for the exotic states at Belle}

\author{C. P. Shen$^*$ for the Belle Collaboration}

\address{Department of Physics, Nagoya University,\\
Nagoya, 4648602, Japan\\
$^*$E-mail: shencp@hepl.phys.nagoya-u.ac.jp\\
www.nagoya-u.ac.jp}

\begin{abstract}

We review recent results on charmonium-like exotic states  from the Belle experiment.
The two-photon process $\gamma \gamma \to  \phi \jpsi$ is measured to
search for $Y(4140)$. No signal for the $Y(4140) \to \phi \jpsi$ is observed. But a
narrow peak with a significance of 3.2$\sigma$ deviations including systematic
uncertainty is observed at 4350.6 MeV/$c^2$ that we named $X(4350)$. We also
search for charmonium-like states, including $X(3872)$, $Y(4140)$, $X(3915)$ and
$X(4350)$, in $\Upsilon(1S)$ and $\Upsilon(2S)$ radiative decays. No significant signal of any
charmonium-like state is observed. The processes $\gamma \gamma \to VV$ ($V=\omega$ or $\phi$) are also measured to
 search for the possible exotic states in low mass region. There are clear
 resonant structures in all the decay modes.

\end{abstract}


\bodymatter

%%%%%%%%%%%%%%%%%%%%%%%%%%%%%%%%%%%%%%%%%%%%%%%%%%%%%%%%%%%%%%%%%%%%%%%%%%%%%%%%%%%%%%%%%%%%%%%%%
\section {Search for the $Y(4140)$ in $\gamma
\gamma \to \phi J/\psi$}

The CDF Collaboration reported evidence of a state called $Y(4140)$
in $B^+ \to K^+ \phi \jpsi$.  No
significant signal at Belle was found although the upper limit on the
production rate does not contradict the CDF measurement.
And also no expected $Y(4140)$ signal was observed
at LHCb experiment~\cite{lhcb}.

The Belle Collaboration searched for this
state in two-photon production~\cite{x4350}
based on a 825~fb$^{-1}$ data sample. No $Y(4140)$ signal is
observed, and the upper limit on the product of the two-photon decay
width and branching fraction of $Y(4140) \to \phi J/\psi$ is
measured to be $\Gamma_{\gamma \gamma}(Y(4140)) {\cal
B}(Y(4140)\to\phi J/\psi)<41~\hbox{eV}$ for $J^P=0^+$, or
$<6.0~\hbox{eV}$ for $J^P=2^+$ at the 90\% C.L. for the first time.
The upper limit on $\Gamma_{\gamma \gamma}(Y(4140)) {\cal
B}(Y(4140)\to\phi J/\psi)$ from this experiment is lower than the
prediction of $176^{+137}_{-93}$~eV for $J^{PC}=0^{++}$, or
$189^{+147}_{-100}$~eV for $J^{PC}=2^{++}$ (calculated by using the
numbers in Ref.~\cite{tanja}). This disfavors the scenario of the
$Y(4140)$ being a $D_{s}^{\ast+} {D}_{s}^{\ast-}$ molecule with
$J^{PC}=0^{++}$ or $2^{++}$.

\begin{figure}[htbp]
\begin{center}
\psfig{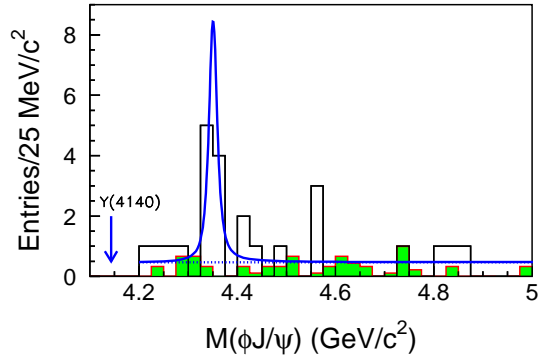}\caption{The $\phi \jpsi$ invariant
mass distribution of the final candidate events. The open histogram
shows the experimental data.  The solid curve is the best fit, the dashed curve is the
background, and the shaded histogram is from normalized $\phi$ and
$\jpsi$ mass sidebands. } \label{mkkjpsi-fit2}
\end{center}
\end{figure}

Evidence is reported for a narrow structure at $4.35~\hbox{GeV}/c^2$
in the $\phi J/\psi$ mass spectrum in
$\gamma \gamma \to \phi J/\psi$ (see Fig.~\ref{mkkjpsi-fit2}).
From the fit, a signal of $8.8^{+4.2}_{-3.2}$ events, with
statistical significance of 3.2$\sigma$, is observed. The mass and natural width
 (named $X(4350)$) are measured to be
$4350.6^{+4.6}_{-5.1}\pm 0.7~\hbox{MeV}$ and $13.3^{+18}_{-9}\pm
4.1~\hbox{MeV}$, respectively. The products of its two-photon decay
width and branching fraction to $\phi J/\psi$ is measured to be
$\Gamma_{\gamma \gamma}(X(4350)) B(X(4350)\to\phi
J/\psi)=6.4^{+3.1}_{-2.3}\pm 1.1~\hbox{eV}$ for $J^P=0^+$, or
$1.5^{+0.7}_{-0.5}\pm 0.3~\hbox{eV}$ for $J^P=2^+$. It is noted that
the mass of this structure is consistent with the predicted values
of a $c\bar{c}s\bar{s}$ tetraquark state with $J^{PC}=2^{++}$ in
Ref.~\cite{stancu} and a $D^{\ast+}_s {D}^{\ast-}_{s0}$ molecular
state in Ref.~\cite{zhangjr}.

%%%%%%%%%%%%%%%%%%%%%%%%%%%%%%%%%%%%%%%%%%%%%%%%%%%%%%%%%%%%%%%%%%%%%%%%%%%%%%%%%%%%%%%%%%%%%%%%%%%%
\section{Search for charmonium-like states in  $\Upsilon(1S)$ and $\Upsilon(2S)$ radiative decays}

To better understand the so-called ``$XYZ$
particles", it is necessary
to search for such states in more production processes and/or
decay modes. For charge-parity-even
charmonium-like states, radiative decays of the narrow $\Upsilon$
states below the open bottom threshold can be examined.
The data used in such analysis~\cite{y1srad, y2srad} include
102 million $\Upsilon(1S)$ events and 158 million $\Upsilon(2S)$ events.

The $X(3872)$ signal is searched for via $X(3872)\to \pp \jpsi$
and $\pp \pi^0 \jpsi$ processes. Except for a few residual ISR
produced $\psp$ signal
events, only a small number of events appear in the $\pp\jpsi$
invariant mass distributions for both $\Upsilon(1S)$ and $\Upsilon(2S)$ decays. In the
$\pp\pi^0\jpsi$ invariant mass distributions, we
observe two events in the $\Upsilon(1S)$ data with masses of 3.67~GeV/$c^2$ and
4.23~GeV/$c^2$; while a few events in the $\Upsilon(2S)$ data.

We search for the $X(3915)$ in the $\omega\jpsi$
mode. No event is observed within the $X(3915)$ mass region in $\Upsilon(1S)$ data;
while there is one event with $m(\pp \piz
\jpsi)$ at 3.923~GeV/$c^2$ and $m(\pp \piz)$ at 0.790~GeV/$c^2$
from $\Upsilon(2S)$ data.

We also search for the $Y(4140)$ in both $\Upsilon(1S)$ and $\Upsilon(2S)$ data,
and $X(4350)$ in $\Upsilon(2S)$ data only in the $\phi\jpsi$ mode.
Nor are there candidate events in the $Y(4140)$ or $X(4350)$ mass regions.
Table~\ref{summary} lists final results for the upper
limits on the branching fractions of $\Upsilon(1S)$ and $\Upsilon(2S)$
radiative decays.

\begin{table}
\tbl{Summary of the limits on $\Upsilon(1S)$ and  $\Upsilon(2S)$ radiative decays to
charmonium-like states $R$. Here $\BR(\Upsilon \to \gamma R)^{\rm up}$ (${\cal
B}_R$) is the upper limit at the 90\% C.L. on the product
branching fraction in the case of a charmonium-like state.}
{\begin{tabular}{@{}ccc@{}}
\toprule
 State ($R$)& ${\cal B}_R$ ($\Upsilon(1S)$) & ${\cal B}_R$ ($\Upsilon(2S)$) \\
\hline
 $X(3872) \to \pp \jpsi$  & $1.6\times 10^{-6}$ &  $0.8\times 10^{-6}$\\
 $X(3872) \to \pp \pi^0 \jpsi$  & $2.8\times 10^{-6}$ & $2.4\times 10^{-6}$ \\
 $X(3915) \to \omega \jpsi$  & $3.0\times 10^{-6}$ & $2.8\times 10^{-6}$  \\
 $Y(4140) \to \phi \jpsi$ & $2.2\times 10^{-6}$ &$1.2\times 10^{-6}$ \\
 $X(4350) \to \phi \jpsi$ & $\cdots$ & $1.3\times 10^{-6}$ \\\botrule
\end{tabular}}\label{summary}
\end{table}

%%%%%%%%%%%%%%%%%%%%%%%%%%%%%%%%%%%%%%%%%%%%%%%%%%%%%%%%%%%%%%%%%%%%%%%%%%%%
\section {Observation of new resonant structures in $\gamma \gamma \to \omega \phi$, $\phi
\phi$ and $\omega \omega$}

Recently in the two-photon processes
$\gamma\gamma\to \omega\jpsi$ and $\phi
\jpsi$, a state $X(3915)$~\cite{x3915} and an evidence for
$X(4350)$~\cite{x4350} were observed, respectively.
It is natural to extend the above theoretical picture to similar
states coupling to $\op$, since the only difference between such
states and the $X(3915)$~\cite{x3915} or $X(4350)$~\cite{x4350} is
the replacement of the $c\bar{c}$ pair with a pair of light quarks.
States coupling to $\oo$ or $\phi\phi$ could also provide information on the
classification of the low-lying states coupled to pairs of light vector
mesons.

\begin{figure}
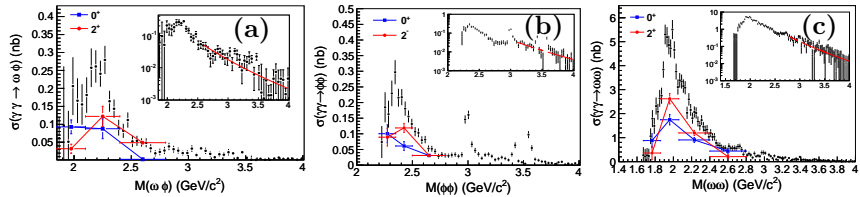

\begin{center}
\psfig{file=fig2a.epsi,angle=-90,width=1.55in}
\psfig{file=fig2b.epsi,angle=-90,width=1.4in}
\psfig{file=fig2c.epsi,angle=-90,width=1.4in}
 \put(-240,-12){ \bf (a)}
 \put(-130,-11){ \bf (b)}
 \put(-25,-11){ \bf (c)}
 \end{center}
\caption{The cross sections of $\gamma \gamma \to \omega \phi$
(a), $\phi \phi$ (b), and $\omega \omega$ (c)
are shown as points with error
bars. The fraction contributions for different $J^P$ values as a
function of $M(VV)$  are shown as the points and squares with error bars.} \label{cross-section}
\end{figure}

The $\gamma \gamma \to VV$ cross sections are shown in
Fig.~\ref{cross-section}~\cite{gg2vv}.
The fraction of cross sections for
different $J^P$ values as a function of $M(VV)$ is also shown in
Fig.~\ref{cross-section}. We conclude that there are at least two
different $J^P$ components ($J=0$ and $J=2$) in each of the three
final states. The inset also shows the distribution of the cross
section on a semi-logarithmic scale, where, in the high
energy region, we fit the $W^{-n}_{\gamma \gamma}$ dependence of
the cross section.

We observe
clear structures at $M(\op)\sim 2.2$~GeV/$c^2$, $M(\phi \phi)\sim 2.35$~GeV/$c^2$,
and $M(\oo)\sim 2.0$~GeV/$c^2$. While there are substantial spin-zero components in
all three modes, there are also spin-two components near threshold.

%%%%%%%%%%%%%%%%%%%%%%%%%%%%%%%%%%%%%%%%%%%%%%%%
\section{Summary}

We review some results on exotic states search from the Belle experiment
in two-photon processes $\gamma \gamma \to  \phi \jpsi$,  $VV$ ($V=\omega$ or $\phi$)
and also in $\Upsilon(1S)$ and $\Upsilon(2S)$ radiative decays.

%%%%%%%%%%%%%%%%%%%%%%%%%%%%%%%%%%%%%%%%%%%%%%%%%%%%%%%%%%
\section*{Acknowledgments}

This work is partially supported by a Grant-in-Aid for Scientific
Research on Innovative Areas ``Elucidation of New Hadrons with a
Variety of Flavors'' from the ministry of Education, Culture,
Sports, Science and Technology of Japan.

%%%%%%%%%%%%%%%%%%%%%%%%%%%%%%%%%%%%%%%%%%%%%%%%%%%%%%%%%%%%%%%%%%%%%%%%%%%%%%%%%%%%%%%%%%%%%%%%

%%%%%%%%%%%%%%%%%%%%%%%%%%%%%%%%%%%%%%%%%%%%%
\bibliographystyle{ws-procs9x6}
\bibliography{ws-pro-sample}

\end{document}